\documentclass[a4paper,superscriptaddress,twocolumn,prl]{revtex4-1}
\usepackage{graphicx}
\usepackage{amsmath}
\usepackage{xcolor}
\setcitestyle{super}
\usepackage{siunitx} 
\usepackage{gensymb}

\newcommand{\upcite}[1]{\cite{#1}}
\newcommand{\bs}[1]{\boldsymbol{#1}}

\begin{document}
\title{Observation of Breathers in an Attractive Bose Gas}

\author{P.J. Everitt}	
\email{patrick.everitt@anu.edu.au}
\homepage{http://atomlaser.anu.edu.au/}
\author{M.A. Sooriyabandara}
\author{G.D. McDonald}
\author{K.S. Hardman}
\author{C.Quinlivan}
\author{P.Manju}
\author{P.Wigley}
\author{J.E. Debs}
\author{J.D. Close}
\author{C.C.N. Kuhn}
\author{N.P. Robins}

\affiliation{Quantum Sensors and Atomlaser Lab, Department of Quantum Science, Australian National University, Canberra, 0200, Australia}

\date{\today} 

\maketitle

\textbf{
In weakly nonlinear dispersive systems, \textit{solitons} are spatially localized solutions which propagate without changing shape through a delicate balance between dispersion and self-focusing nonlinear effects\upcite{ablowitz_solitons_1991}. These states have been extensively studied in Bose-Einstein condensates (BECs), where interatomic interactions give rise to such nonlinearities\upcite{kh._abdullaev_dynamics_2005}. Previous experimental work with matter wave solitons has been limited to static intensity profiles. The creation of matter wave \textit{breathers}---dispersionless soliton-like states with collective oscillation frequencies driven by attractive mean-field interactions---have been of theoretical interest\upcite{matuszewski_stability_2005,cardoso_modulation_2010} due to the exotic behaviour of interacting matter wave systems. Here, using an attractively interacting Bose-Einstein condensate, we present the first observation of matter wave breathers. A comparison between experimental data and a cubic-quintic Gross-Pitaevskii equation (GPE) suggests that previously unobserved three-body interactions may play an important role in this system. The observation of long lived stable breathers in an attractively interacting matter wave system indicates that there is a wide range of previously unobserved, but theoretically predicted, effects that are now experimentally accessible\upcite{trombettoni_discrete_2001}. A compelling example is the recently predicted quantised tunnelling of a breather through a barrier\upcite{dunjko_superheated_2014}.}\\ 

The study of solitons has found application in fields as diverse as nonlinear optics\upcite{kibler_peregrine_2010}, magnetic materials\upcite{kosevich_magnetic_1998}, and atom interferometry\upcite{mcdonald_bright_2014}. Both dark\cite{frantzeskakis_dark_2010} and bright matter wave solitons have been realized experimentally and have been shown in exotic contexts such as colliding solitons\upcite{nguyen_collisions_2014}, soliton formation during cloud collapse\upcite{cornish_formation_2006}, and the collision of bright solitons with a barrier\upcite{marchant_controlled_2013}. 

Breathers are thought to be a similarly ubiquitous phenomenon\cite{kivshar_dynamics_1989}. They have been experimentally realised in a variety of systems including Josephson-junction lattices, spin lattices, and optical waveguide arrays\upcite{flach_discrete_2008,mandelik_observation_2003}. Breathers are also thought to play a role in more exotic contexts such as DNA denaturation\upcite{peyrard_using_1998}. While breathers have been predicted to exist in matter wave systems\upcite{trombettoni_discrete_2001}, they have not been observed until now.

 \begin{figure}[!hbp]
 	\centering{}
 	\includegraphics[width=1\columnwidth]{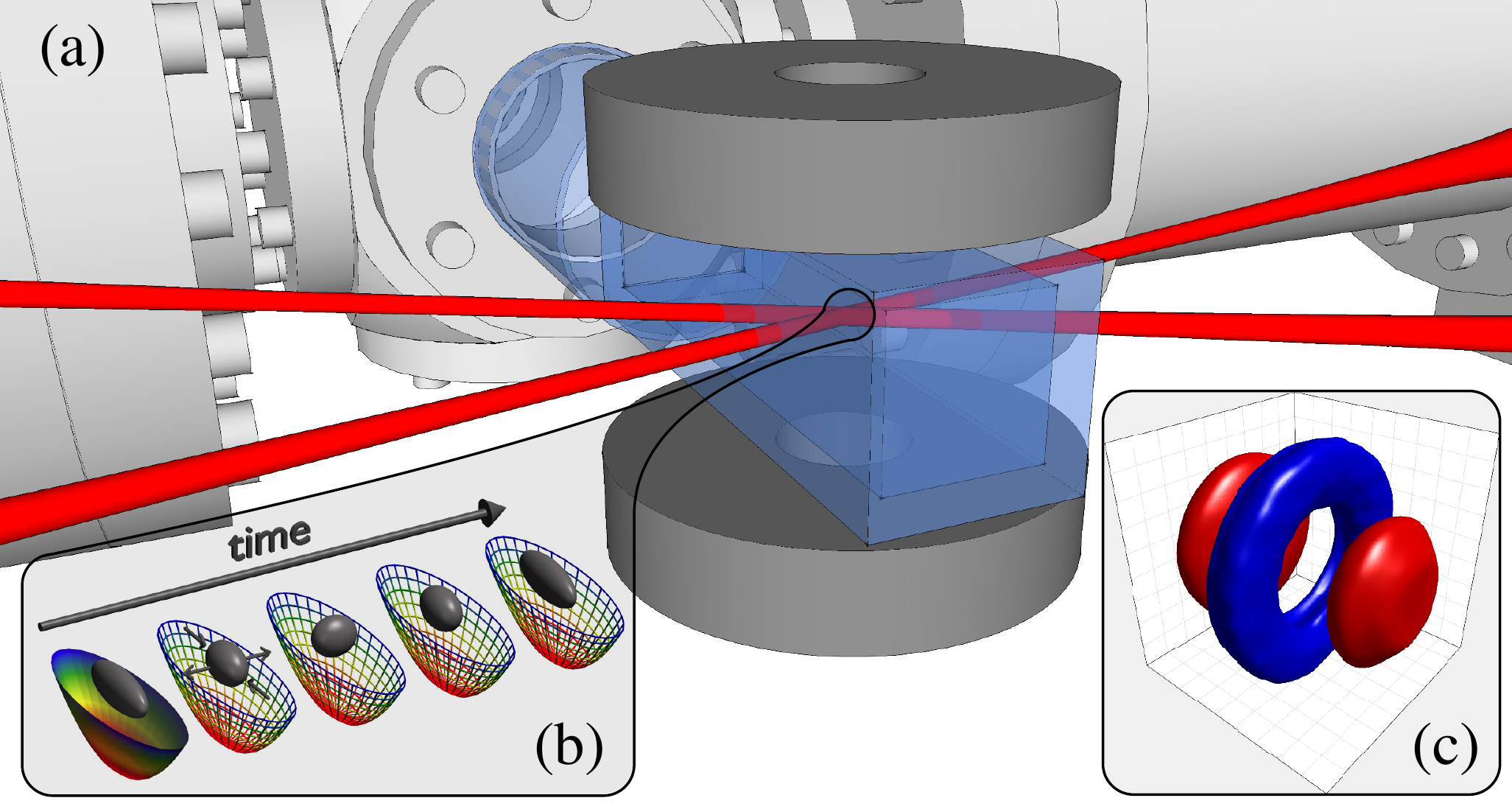}
 	\caption{a) The UHV apparatus, with the glass cell centred. Solenoids top and bottom provide a magnetic bias field of $\sim$155G while intersecting laser beams at 1064nm generate a potential that traps the atoms. b) Schematic of a BEC undergoing a quadrupole oscillation inside a cigar-shaped optical dipole trap. The quadrupole oscillation occurs around the weaker axial frequency of the harmonic optical trapping potential holding the atoms. c) The quadrupole collective normal mode. Red and blue isosurfaces are out of phase contributions to the condensate density oscillations. The mode was tomographically reconstructed from dual axis imaging data using principal component analysis and an inverse radon transform with imposed cylindrical symmetry. }
 	\label{fig:schematic}
 \end{figure}
 
To create the matter wave breathers, an attractively interacting Bose-condensed cloud of $^{85}$Rb atoms is made to undergo collective quadrupole oscillations inside an anisotropic cigar-shaped trap (see Figure \ref{fig:schematic}).  A sequence of measurements are made on the excitations of trapped clouds at varying s-wave scattering lengths, $a_s$, the critical parameter determining the lowest order interactions of the atoms. In the regime where $a_s<0$, stable quadrupole oscillations above the frequency of those measured at $a_S=0$ are observed, indicating that the quadrupole mode in this regime is being driven by the attractive s-wave interactions, confirming the existence of breathers \upcite{Akhmediev}. 

In addition to the observation of matter wave breathers, we find quantitative evidence of physics not captured by two-body scattering alone. Firstly, a measurement of the quadrupole oscillation frequency against the fundamental centre of mass (CoM) mode of the trap in the absence of two-body interactions ($a_S=0$) gives an unexpected, but previously predicted ratio. Secondly, at negative s-wave scattering lengths ($a_s<0$) the trapped condensate is also observed to be stable over a larger range of scattering lengths than previously predicted. Finally, an anomalous phase shift is observed between the axial and radial quadrupole oscillations in expansion data. It is shown that the presence of a repulsive three-body interaction quantitatively explains the observations. 
 
Initially, the quadrupole mode oscillation was probed in the absence of two-body interactions ($a_{s}=0$). The scheme described in Methods was used to excite the quadrupole mode ($\omega_Q$) at $a_{s}=0$ and the oscillation frequency was measured by fitting the averaged widths to a decaying sine curve. This was compared to a direct measurement of the axial trapping frequency ($\omega_{z}$) made by using Bragg transitions to excite the centre of mass (CoM) mode. Data was acquired by alternating between quadrupole and CoM measurements to reduce the chance of systematic drift influencing the measured ratio $\omega_Q/\omega_z$. It was found that at $a_{s}=0$, $\omega_{Q}=(1.67\pm 0.03) \omega_{z}$ with $\omega_{z}=2\pi \times(7.55\pm0.15) \si{\hertz}$.  This ratio differs significantly from the noninteracting gas ($a_{S}=0$) result which predicts a ratio of 2 between quadrupole and CoM modes. Measurements made at positive scattering length for both the $^{85}$Rb BEC and a $^{87}$Rb BEC in the same trap give results consistent with theoretical predictions ($\omega_{Q}=\sqrt{5/2} \omega_{z}$) \upcite{pollack_collective_2010}.  $^{85}$Rb is observed to have a significant three-body loss around the Feshbach resonance at 155G \cite{altin_collapse_2011} and, intriguingly, it is predicted that there should also be significant three-body scattering \cite{kohler_three-body_2002}. The ratio measured here for the noninteracting gas is in excellent agreement with a theoretically predicted frequency ratio which accounts for non-zero three-body interactions at $a_{s}=0$\cite{al-jibbouri_geometric_2013} (see supplementary material). 

 \begin{figure}[!htp]
 	\centering{}
 	\includegraphics[width=1\columnwidth]{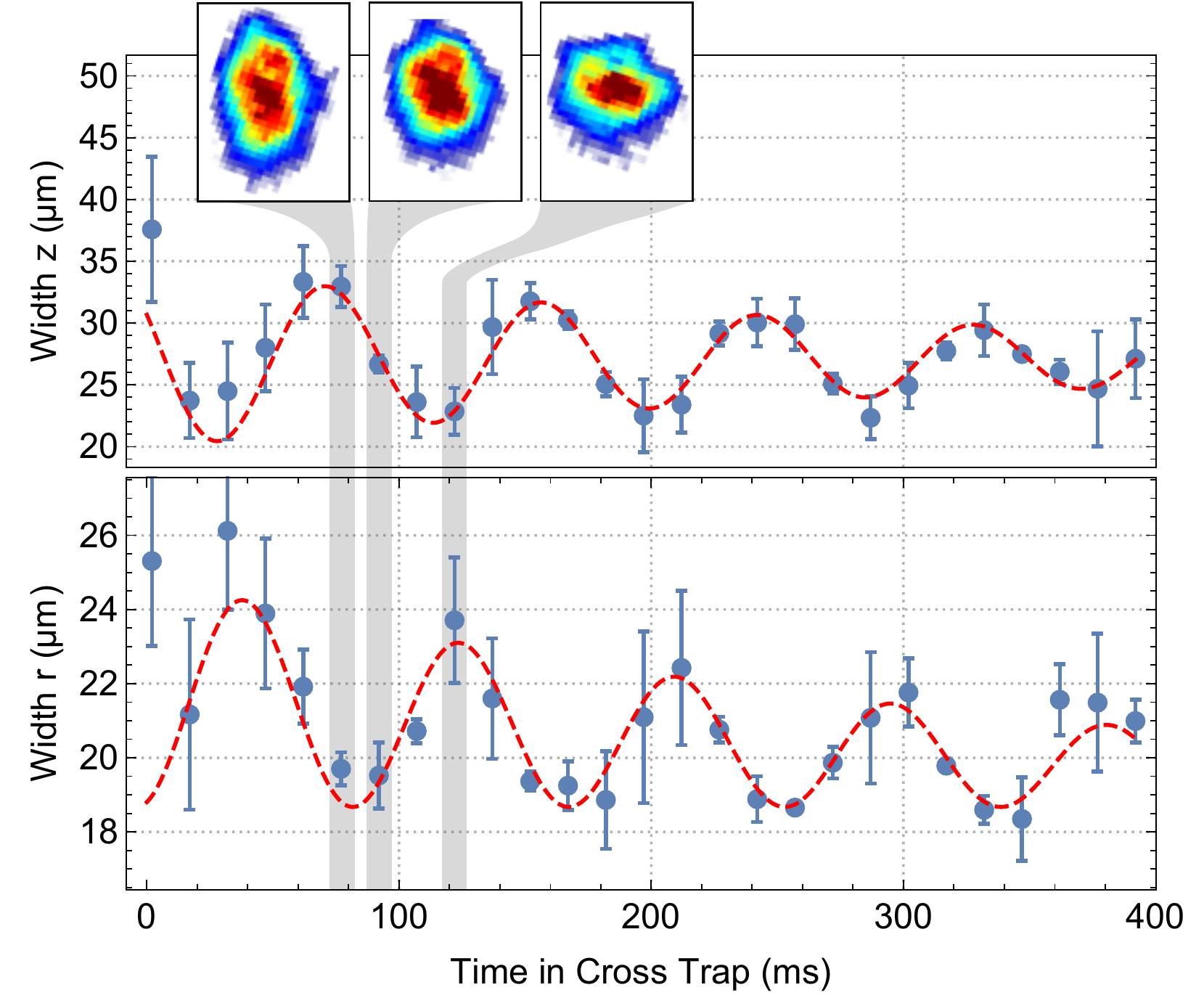}
 	\caption{The measured radial (r) and axial (z) widths of the oscillating cloud in the cross trap after the scattering length was jumped to $-13 a_{0}$. Each point is a weighted average of multiple shots ($\sim$4). Fitted decaying sine curves with $\omega_{Q}=2\pi\times \left(11.54 \pm 0.12 \right)$ \si{\hertz} are also plotted. Inset are absorption images at \SI{77}{\milli\second}, \SI{92}{\milli\second} and \SI{122}{\milli\second} showing the cloud changing from maximum to minimal axial extension. The radial and axial widths oscillate almost out of phase, with a small phase shift of $\left(0.24\pm0.06\right)\pi$ due the free space expansion of the condensate prior to imaging (see supplementary material).}
 	\label{fig:oscillations}
 \end{figure}

Having investigated the $a_s=0$ case, the quadrupole mode oscillation was excited at varying {\em negative} scattering lengths, $a_s<0$. In these experiments, the trapped cloud of $\sim$$10^4$ atoms is observed to be stable against condensate collapse over a broader range than expected from a mean field theory with only two-body scattering (see supplementary material for discussion and data). In this regime, the negative s-wave scattering length will significantly modify the collective excitation spectrum.  Figure \ref{fig:oscillations} shows the axial and radial oscillations of an N=$1.5\times10^{4}$ BEC, induced by jumping the scattering length from $a_s=40a_0$ to $a_s=-13a_0$, where $a_0$ is the Bohr radius. The oscillations are stable for over \SI{400}{\milli\second}. For $a_s = -13a_0$ the quadrupole mode frequency is determined to be $\omega_{Q}=2\pi\times \left(11.54 \pm 0.12 \right)$ \si{\hertz}. A small phase shift of $\left(0.24\pm0.06\right)\pi$ between the axial and radial widths is present. This shift arises during the 20ms of free space expansion before absorption imaging of the cloud. In the supplementary material, it is shown that this shift during the expansion is quantitatively predicted by an extended GPE containing a quintic term\cite{kohler_three-body_2002}:

\begin{align}
i \hbar \frac{\partial \Psi(\bs{r},t)}{\partial t} &= \left( -\frac{\hbar^{2}}{2m} \nabla^{2} + V(\bs{r})  + g_{2} N |\Psi(\bs{r},t)|^{2}\right. \nonumber \\
&+ g_{3}  N^2 |\Psi(\bs{r},t)|^{4} \biggr) \Psi(\bs{r},t) 
\label{eq:GP_equation}
\end{align} 

where $\bs{r} = (x,y,z)$ is a position vector, $N$ is the condensate population, $m$ is the mass of an $^{85}$Rb atom, $V(\bs{r})$ is the trapping potential, $g_{2}=\frac{4 \pi \hbar^{2} a_{s}}{m}$ is the two-body interaction term, and $g_3$ is the three-body interaction term. For a cylindrical cigar-shaped trap, as in our system, the potential is given by $V(\bs{r})=\frac{m}{2} \left( \omega_{z} z^{2}+ \omega_{r} (x^2+y^2)\right)$. 

\begin{figure}[!htp]
	\centering{}
	\includegraphics[width=1\columnwidth]{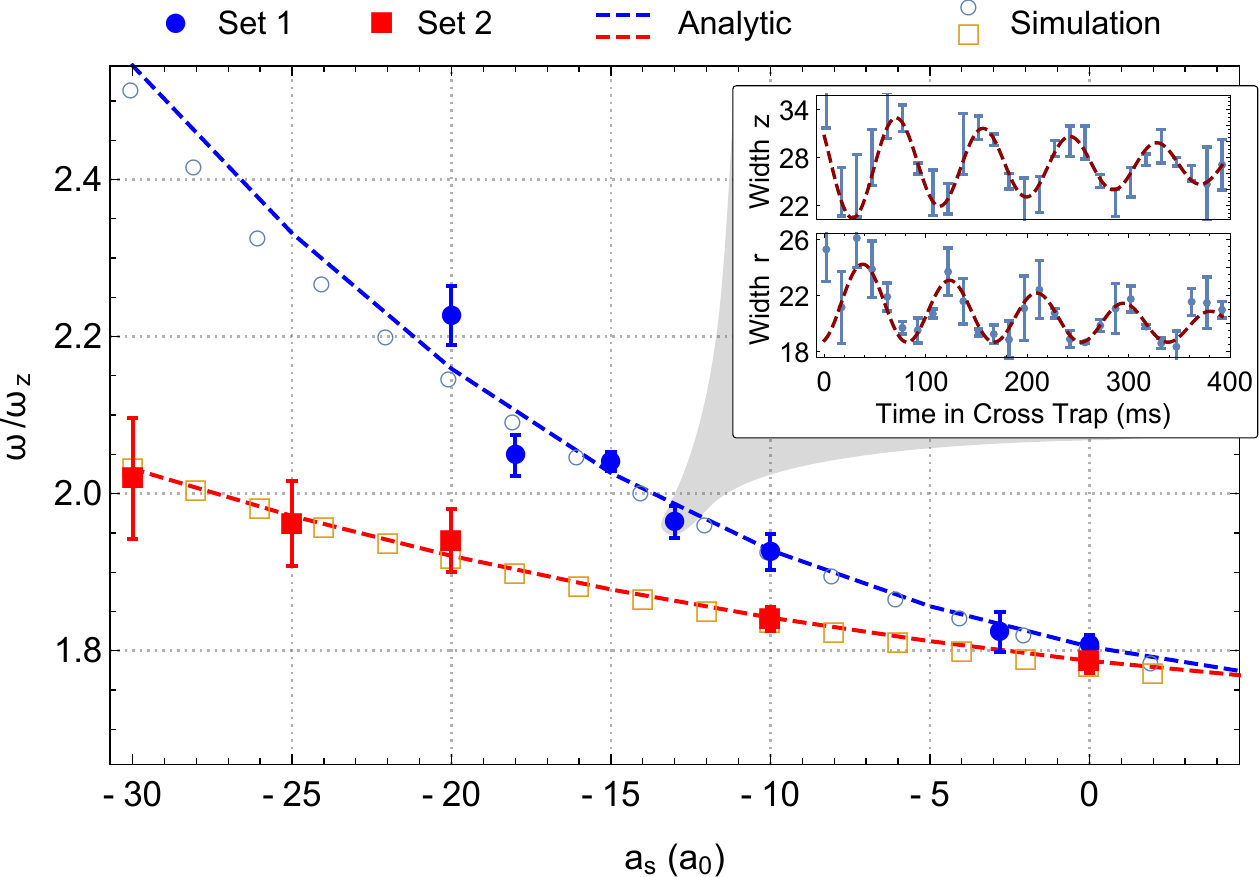}
	\caption{The measured oscillation frequency against the scattering length for set 1 ($N=1.5\times10^{4}$) and set 2 ($N=1\times10^{4}$). Frequencies are normalised by $\omega_{z}$ which is calculated from the interactionless ($a_{s}$=0) oscillations. Dashed lines indicate the theoretical frequencies of equation \eqref{eq:frequencies} using parameters \{N=$1.5\times10^{4}$, $\omega_{z}=2\pi\times5.88$  \si{\hertz}, $\omega_{r}=2\pi\times77$  \si{\hertz}, $\text{Re}[g_{3}]=1.23\times10^{-25}\hbar\si{cm^6s^{-1}}$\} and \{N=$1.0 \times10^{4}$, $\omega_{z}=2\pi\times6.61$  \si{\hertz}, $\omega_{r}=2\pi\times77$  \si{\hertz}, $\text{Re}[g_{3}]=3.3\times10^{-25}\:\hbar \si{cm^6s^{-1}}$\} for sets 1 and 2 respectively. Open data points are simulated frequencies for the same parameters. The inset figure demonstrates the oscillation measurement taken for each data point.}
	\label{fig:soliton_frequencies}
\end{figure}

By conducting measurements from $a_s=0$ through a range of negative scattering lengths the dependence of the quadrupole frequency on $a_s$ can be determined (Figure \ref{fig:soliton_frequencies}). The results from two separate experiments (the optical trap was rebuilt between data sets) using condensed populations of $N=1.5\times10^{4}$ and $N=1\times10^{4}$ are plotted. The data for different $a_{s}$ were taken out of order to remove systematic effects on the measured frequencies. As $a_{s}$ becomes more negative the oscillation frequency becomes greater than the measured interactionless ($a_s=0$) quadrupole mode. An observation of a collective oscillation frequency {\em higher} than the measured interactionless case is the hallmark of a breather.  

An analytic expression for the quadrupole excitation frequency for a cubic-quintic GPE has previously been derived \upcite{al-jibbouri_geometric_2013}:

\begin{align}
\omega_{Q}=& \biggl(2+4k J_{6,2}+2 \lambda^{2} -2P J_{2,3}  \nonumber \\
-2& \sqrt{  \left( 1+2k J_{6,2}-\lambda^{2}+P J_{2,3} \right)^{2} +8 \left( P J_{3,2} +2k J_{5,3}\right)^2} \biggr)
\label{eq:frequencies}
\end{align}

where $\lambda = \omega_{z}/\omega_{r}$ is the trap aspect ratio, $P=\frac{\sqrt{2/\pi} N a_{s}}{\ell}$ is the dimensionless two-body interaction parameter, and $k=\frac{4\textrm{Re}[g_{3}] N^2}{9\sqrt{3} \pi^{3} \hbar \omega_{r} \ell^{6}}$ is a dimensionless three-body interaction parameter. $J_{i,j}=1/(4u_{r0}^{i}u_{z0}^{j})$, where, respectively, $u_{r0}$ and $u_{z0}$ are the equilibrium axial and radial widths of the cloud in dimensionless units. 

As highlighted in the supplementary material, a number of independent quantitative pieces of evidence point to a three-body interaction parameter playing a role in the system dynamics. In the following, the experimental data from Figure \ref{fig:soliton_frequencies} are analysed {\em assuming} that there is indeed such a three-body interaction. 
To compare with theory, equation \eqref{eq:frequencies} was fitted to the experimental data by varying the value of $g_{3}$, with $\omega_{z}$ being set by the measured interactionless oscillation. Figure \ref{fig:soliton_frequencies} shows the resulting fit for the two sets of data. Fitted values are $\mathrm{Re}[g_{3}]=(1.23\pm 0.09\pm 0.2)\times10^{-25}\:\hbar \si{cm^6s^{-1}}$ and $(3.3\pm 0.4\pm1.0)\times10^{-25}\:\hbar \si{cm^6s^{-1}}$ for sets 1 and 2 respectively, well within the range predicted by theory \upcite{hao-cai_boseeinstein_2010,gammal_critical_2001,pieri_derivation_2003} (see further discussion in the supplementary materials). The first uncertainty quoted for $\mathrm{Re}[g_{3}]$ is statistical while the second is from systematic effects such as atom number variation. Quadrupole oscillation frequencies obtained from numerical simulations of Equation \ref{eq:GP_equation} are also presented in Figure \ref{fig:soliton_frequencies}, showing excellent agreement with the analytical model.  For the data in Figure \ref{fig:soliton_frequencies}, the ratio predicted by equation \ref{eq:frequencies} at $a_s=0$ is $\omega_{Q}/\omega_{z}=1.8$ for both sets. Equation \ref{eq:frequencies} also predicts the measured ratio of CoM to quadrupole oscillations from measurements with Bragg ($\omega_{Q}/\omega_{z}=1.7$ for the experimental conditions in that measurement). 

All axial oscillations in the range $a_s=[-30a_{0},0a_{0}]$ are seen to occur around a stable mean width (up to a small decay from atom loss), as in Figure \ref{fig:oscillations}. It is also noted that as $a_{s}$ decreases the quadrupole mode frequency becomes higher than the $a_{s}=0$ case indicating that the oscillations are driven by nonlinear self-interaction of the cloud rather than the trapping potential alone \upcite{Akhmediev,PhysRevLett.30.1262}.
 
In conclusion, the first observation of a matter wave breather driven by attractive two-body interactions was made. The dynamics of these breathers were found to agree with the predictions of a cubic-quintic GPE equation with repulsive three-body interactions. The extended GPE was found to closely model the stability of the condensate at negative scattering length, the measured values of the breather oscillation frequencies, as well as the phase shift due to cloud expansion, yielding an estimate for the three-body scattering parameter $g_3$ in $^{85}$Rb to be of the order $10^{-25}\hbar$\si{cm^6s^{-1}}. 
 
It is clear from the data presented here that further experiments aimed at elucidating the three-body interaction are required.  Although this higher order interaction quantitatively models the experiment very well, a key result indicates that there is more to the story: when the two sets of data presented in Figure \ref{fig:soliton_frequencies} are used to constrain both the analytic and numerical models, different values for the $g_3$ parameter are found. The results from each set are different by a factor of roughly two and is outside our statistical and known systematic uncertainties. A re-examination of previous experiments in the attractive two-body regime for $^{85}$Rb, such as the Bosenova \upcite{donley_dynamics_2001} and soliton formation \upcite{cornish_formation_2006}, may provide additional insights. Furthermore, it may be the case that higher order interactions are observable in other BECs of other species. Indeed, evidence of such interactions has recently been found in exciton-polariton condensates\cite{snoke2015}.
 
Work is ongoing in a number of directions.  First, to extend this study to freely propagating breather states, by removing the axial trapping potential and releasing the atoms into an optical waveguide \upcite{mcdonald_bright_2014}. The Bragg system used to excite the sloshing mode of the cloud can also be applied to transfer quantised centre of mass momentum to the breather state, as well as splitting the cloud into multiple momentum states with varying ratios\upcite{kozuma_coherent_1999}. Combined with potential barriers created by blue-detuned, tightly focused lasers, we are now in a position to begin studying the exotic nonlinear physics of multiple interacting matter wave solitons.  Immediate possibilities include examining collisions of excited and ground state solitons, the study of breather solitons interacting with potential barriers, and the examination of the stability and breakup of these new states.

\section*{Methods}
\subsection*{Experimental}
The experimental apparatus is described in depth in previous work\upcite{kuhn_bose-condensed_2014}. In summary, a combined 2D and 3D MOT system collects and cools both $^{85}$Rb and $^{87}$Rb atoms. The atoms are loaded into a magnetic trap and undergo RF evaporation before being loaded into an optical cross trap. The cross trap consists of intersecting $\SI{1090}{\nano\meter}$ and $\SI{1064}{\nano\meter}$ lasers with approximate waists of 300$\mu m$ and 250$\mu m$ (half width at $1/e^2$ intensity), respectively. After loading, the magnetic trap coils are switched from anti-Helmholtz to Helmholtz configuration. This allows the $s$-wave scattering length of the cloud, $a_s$, to be tuned using a Feshbach resonance. Setting the $^{85}$Rb scattering length to zero ($a_{s}=0$) while ramping down the cross trap intensity allows sympathetic cooling to remove the remaining $^{87}$Rb atoms while minimising three-body recombination losses in $^{85}$Rb. A further period of evaporation with $a_{s}=300a_{0}$ creates a pure $^{85}$Rb BEC with atom number of the order of $10^{4}$. This is the initial condition for all experiments. Two orthogonal absorption beams, one of which is co-axial with the vertical 3D MOT beam, allow the cloud to be imaged after 20ms of ballistic expansion ($a_{s}=0$) in free space.

The method of inducing oscillations in the condensate is as follows. The initial BEC is held in the cross trap while the scattering length is adiabatically ramped to a value above that at which the collective oscillations are to be measured (typically $\sim 40a_{0}$). The scattering length is then rapidly jumped (in $<100$\si{\micro\second}) to the desired value. Because the spatial width of the ground state at $a_s = 40$ is much larger than that at $a_s<0$, the cloud begins to oscillate. The trap is then turned off and the cloud allowed to expand at $a_{s}=0$ before an absorption image is taken. By holding the condensate in the cross-trap for different times, the temporal dependence of the cloud width can be measured. Multiple runs of the experiment are taken for each time step to build up statistics. For each absorption image, a 2D Gaussian is fitted to the atomic profile. The radial and axial widths, peak intensity and position are extracted from multiple shots at each time step and an average is constructed by using the error in each fit as weights.  

\subsection*{Simulation}
Simulations were conducted to complement the analytic approach. Equation \eqref{eq:GP_equation} is numerically solved using a 4th-order Runge-Kutta symmetrised split-step Fourier method. The dimension of the simulations is reduced by exploiting the cylindrical symmetry of the condensate to solve the 3D system in 2D cylindrical co-ordinates by a Hankel transform\upcite{ronen_bogoliubov_2006}. 

\section{Acknowledgements}
The authors gratefully acknowledge the support of the Australian Research Council Discovery program. We would also like to thank Maxim Olchanyi and Nail Akhmediev for helpful communications, and Eric Cornell and Randy Hulet for detailed discussions at ICOLS2015.  J.E. Debs would like to acknowledge financial support from the IC postdoctoral fellowship program.

\newpage
\section{Supplementary Material}
\subsection{I. Extracting oscillation frequencies}

\begin{figure}[!htp]
	\centering{}	
	\includegraphics[width=1\columnwidth]{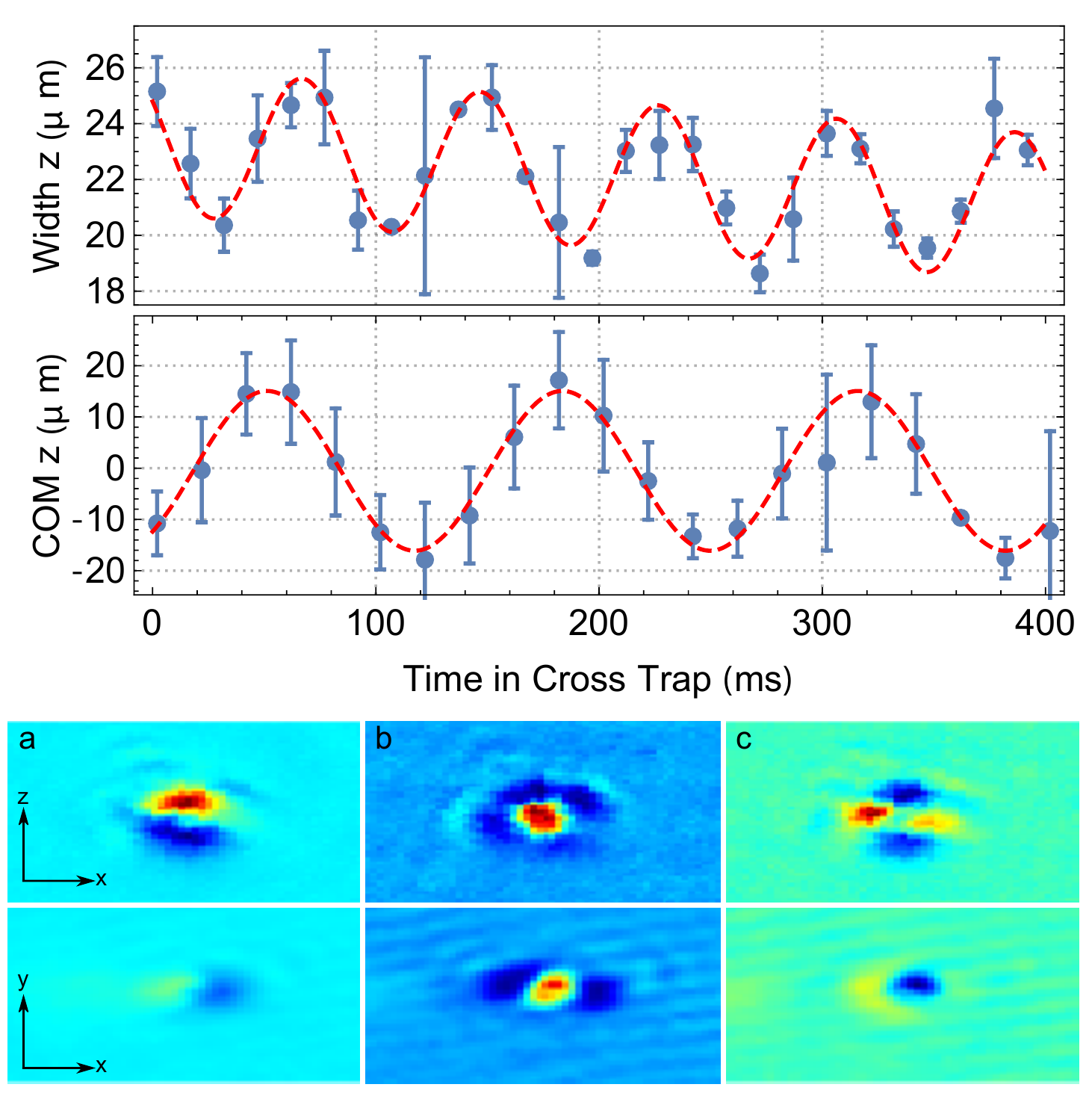}
	\caption{ (sup)  This data was obtained as follows: from BEC creation a 200ms ramp to $40a_0$ in the optical cross trap. The scattering length is then jumped to $0a_0$. The cloud was moved away from the center of the trap using two $2\hbar k$ Bragg pulses 0.7ms apart. Images were then taken in 1ms time steps and the cloud shape observed. Top panels show experimental data for both the CoM oscillation and the quadrupole mode at $a_{s}=0$. The ratio of these frequencies is given in the text. The lower absorption images show the outcome of PCA. The first three principal components (PC) are shown where z is axial, y is vertical, and x is horizontal. a)  Axial CoM mode: vertical oscillations in the upper picture indicate the axial CoM mode. The small signal in the lower image is due to the small angle the second camera makes with the waveguide, b) Monopole Mode: in phase axial and radial oscillations, and c)  Quadrupole Mode: out of phase axial and radial oscillations. }
	\label{fig:CoM}		
\end{figure}

The observation of trap oscillations frequencies for $a_s \le 0$ in $^{85}$Rb required extreme stability of the apparatus. Condensates with atom number variation at the 5\% level over 12 hours and low noise, multi-directional imaging were required to obtain a strong signal.  All data sets were collected non-sequentially and different data sets were taken on multiple days and under different experimental conditions.  The Feshbach magnetic field is characterised to significantly below an uncertainty of $1\,$mG using radio frequency spectroscopy of the trapped cloud, resulting in a precise, absolute knowledge of $a_s$ at the sub $0.1a_0$ level \cite{mcdonald_bright_2014}. 
The conclusions of the paper were drawn from a number of different analysis techniques. Critical to all analysis was the simultaneous absorption imaging of the atom cloud in two orthogonal directions, one with an angle of $11.7\degree$ with the axial direction and one perpendicular.  The orthogonal images obtained allowed the identification and measurement of all the lower order oscillations of the trapped cloud.  The top panels of Figure \ref{fig:CoM}(sup) shows oscillation data extracted by simply fitting a Gaussian profile to the absorption image of the cloud after 20ms of expansion at zero scattering length.  The lower panels of Figure \ref{fig:CoM}(sup) shows the outcome of principal component analysis\upcite{1367-2630-16-12-122001}(PCA) on the absorption images combined from the two orthogonal directions: one vertical imaging both the axial and horizontal radial directions (upper image, axial direction is vertical), the other imaging down the waveguide giving a radial cross section (lower image). PCA supplements a more traditional analysis and can recover good approximations to the shape and frequency normal modes of a BEC from a time-series of absorption images.
 
Fitting directly to the cloud and PCA give consistent results regarding oscillation frequencies.  
 
\subsection{II. Oscilations in the presence of three-body interactions}

A variational method has previously been used to calculate the collective excitation spectrum for a BEC in an anisotropic trap\upcite{perez-garcia_low_1996}. This method agrees well with experimental results at positive scattering lengths\upcite{pollack_collective_2010}. For $a_{s}<0$ this model predicts rapid increase of the oscillation frequency and then collapse of the cloud before $a_{s}\approx-2a_{0}$ (for the trap frequencies and atom numbers used in these experiments).  However, the condensates observed here are stable at negative scattering lengths far beyond this prediction, indicating that there is physics not captured by this model which contributes to the condensates' stability. Attractive mean-field interactions will cause the formation of high density regions during collapse where higher order interactions will become important. At large negative scattering lengths, where an attractive two-body interaction would otherwise cause the condensate to catastrophically collapse, a repulsive three-body interaction will stabilise it.

The three-body term is a complex number with Re[$g_{3}$] describing the three body scattering parameter and Im[$g_{3}$] describing the three-body recombination loss. The study of trap loss rates as well as the dynamics of cloud collapse have so far focused on Im[$g_{3}$] with experimental values ranging from $10^{-30}$ to $10^{-25} \:\hbar \si{cm^6s^{-1}}$ for $^{85}$Rb\upcite{bao_three-dimensional_2004,saito_mean-field_2002,altin_collapse_2011,roberts_magnetic_2000}. Experimental quantification of Re[$g_3]$, however, has largely been neglected. Theoretical predictions\upcite{hao-cai_boseeinstein_2010,gammal_critical_2001,pieri_derivation_2003} have ranged from $10^{-27} \:\hbar \si{cm^6s^{-1}}$ to $10^{-24} \:\hbar \si{cm^6s^{-1}}$. \\

By comparing atom number decay curves from experiments and simulations an estimate of the imaginary part of $g_3$ was obtained, $\mathrm{Im}[g_3] = (6\pm2)\times10^{-28}\:\hbar \si{cm^6s^{-1}}$, which is well within the range of published values\upcite{bao_three-dimensional_2004,saito_mean-field_2002,altin_collapse_2011,roberts_magnetic_2000}. \\

\subsection{III. Oscillations of a noninteracting cloud}
The expected frequency of the quadrupole mode when $a_{s}=0$ and $g_{3}=0$ is $\omega_{Q}=2\omega_{z}\;$\upcite{stringari_collective_1996}. However for non-vanishing $g_{3}$, both Equation 2 and numerical results from Equation 1 show that this frequency is lowered by as much as 15\%, depending on the system parameters. While the quadrupole mode frequency is reduced, simulations showed that centre of mass (CoM) oscillations (expected to be at $\omega_{z}$) remained unchanged. A repulsive 3-body interaction term at $a_{s}=0$ predicts a ratio ${\omega_Q/\omega_z < 2}$. The measured ratio at $a_s = 0$ is ${\omega_Q/\omega_z = 1.67\pm0.03}$ with $\omega_{z}=2\pi \times(7.55\pm0.15) \si{\hertz}$, is consistent with Equation 2, for our system parameters and $\text{Re}[g_{3}]=3.3\times10^{-25}\:\hbar \si{cm^6s^{-1}}$. In Figure 1(sup) the experimental data for the quadrupole mode is shown compared to the measurement of the CoM. It has previously been suggested that the two body interactionless regime ($a_{s}=0$) is the ideal place to search for additional 3-body physics \upcite{gammal_critical_2001}.  Thus, this measurement alone constitutes an observation that there is significant 3-body scattering in $^{85}$Rb near the Feshbach resonance at 155G.  Such a shift has {\em not} been observed for oscillations of a Li condensate in the same regime \upcite{pollack_collective_2010, Hulet2015}. While Li and Rb systems are similar, the details of the scattering potentials and atomic structure are critical in determining the interactions for any particular atom \upcite{RevModPhys.82.1225}. As such an absence of three-body scattering in Li does not preclude its presence in Rb. It should also be noted that no attempt has previously been made to measure the excitation spectrum of a non-interacting gas of $^{85}$Rb.\\
\begin{figure}[!htp]
	\centering{}	
	\includegraphics[width=1\columnwidth]{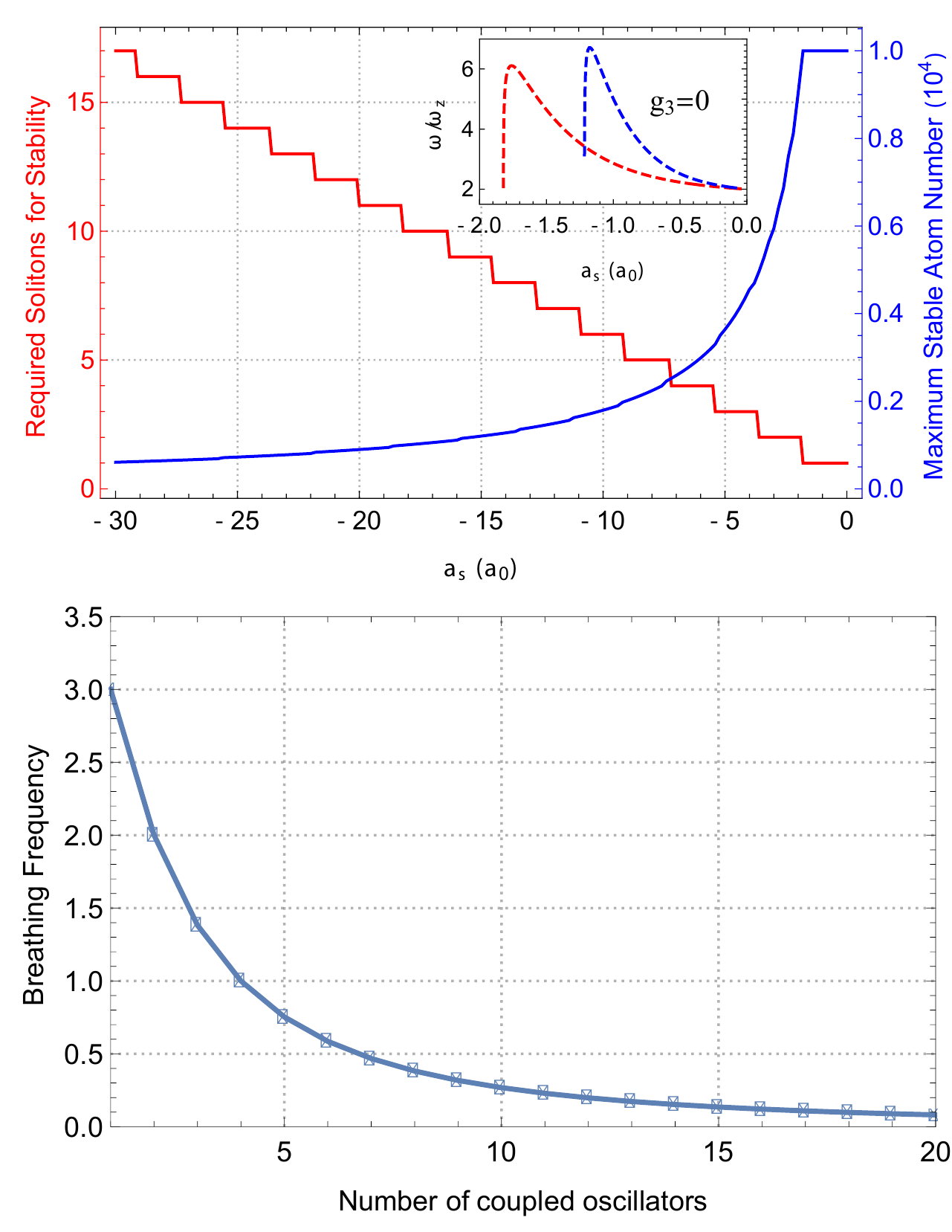}
	\caption{(sup) Upper frame: The maximum predicted stable atom number, and number of solitons that would be produced in the trap used in the experiment, when Re[$g_3$]=0 for negative scattering lengths. Inset is the expected quadrupole oscillation frequency when Re[$g_3$]=0 for the system parameters of both sets in Figure 3(main). Lower frame: a simple model for coupled 1D oscillators indicates that for increasing number of particles the breathing frequency decreases. }
	\label{fig:collapse}		
\end{figure}

\subsection{IV. Stability against collapse}

Perhaps one of the most intriguing indications that three-body interactions are playing a role in this experiment is the observation of large stable atom clouds at $a_s<0$. Stable attractively interacting condensates are observed to exist beyond the limit predicted by a mean field theory considering only two-body interactions\upcite{savage_bose-einstein_2003}. For parameters of data set 2 in the main text Figure 3, collapse is observed at $\sim\mkern-5mu-30a_0$ and indicated by a significant atom loss as well as an increase in the thermal fraction of atoms in the trap. This collapse point is more than a factor of 10 larger than predicted by two-body scattering alone. Numerical simulations of Equation 1 indicate that a non-zero Re[$g_3$] stabilises the atom cloud, allowing condensates to exist in the range observed here.  
 
The observation of condensates violating the prediction of models incorporating two-body scattering alone is not unprecedented. A paper by the JILA group reported condensates with up to 10 times the atom number predicted by GP theory\upcite{cornish_formation_2006}. The interpretation of that experiment was that solitons were forming in-trap post collapse. The absorption images presented in the paper appeared to support this. Numerical simulation of the conditions of that experiment, including our inferred Re[$g_{3}$] interaction, indicate that violent dynamics occur during the 2ms ballistic expansion, because the cloud is directly released from the trap at $-400a_{0}$. The dynamics give rise to structures that appear similar to solitons.

The top frame of Figure \ref{fig:collapse}(sup) illustrates the results of applying the JILA `multi-solition' model to the conditions here.  At the largest negative scattering length used for the breather, 11 solitons would be required to add up to the observed stable atom number. Furthermore, a simple model of coupled oscillators indicates that a soliton chain breathing mode frequency should dramatically decrease with an increasing number of discrete solitons (see the lower panel of Figure \ref{fig:collapse}(sup)). The observation in this paper is that the breather frequency {\em increases} smoothly.      	

It could be argued that the imaging system used here is simply not able to resolve soliton trains. In order to investigate this hypothesis, further experiments have been conducted in which the condensate was released into the waveguide using a method described previously\upcite{mcdonald_bright_2014}. By carefully choosing initial conditions and s-wave scattering lengths, soliton trains were produced via a modulational instability mechanism in the single beam waveguide.  Figure \ref{fig:trains}(sup) shows that the imaging system is clearly able to resolve large soliton trains. 

Finally, it is highly unlikely that the symmetric monopole mode of the in-trap oscillation extracted by PCA in Figure 1(sup) would arise from a multicomponent soliton. The reasoning is twofold: firstly, it is unlikely that the breathing oscillation of each individual soliton in the strongly trapped direction would be in phase; secondly, it is unlikely that the breathing oscillation of each individual soliton would have the same frequency as the symmetric oscillation of the soliton train in the weakly trapped axis. 

\begin{figure}[!htp]
 	\centering{}	
 	\includegraphics[width=1\columnwidth]{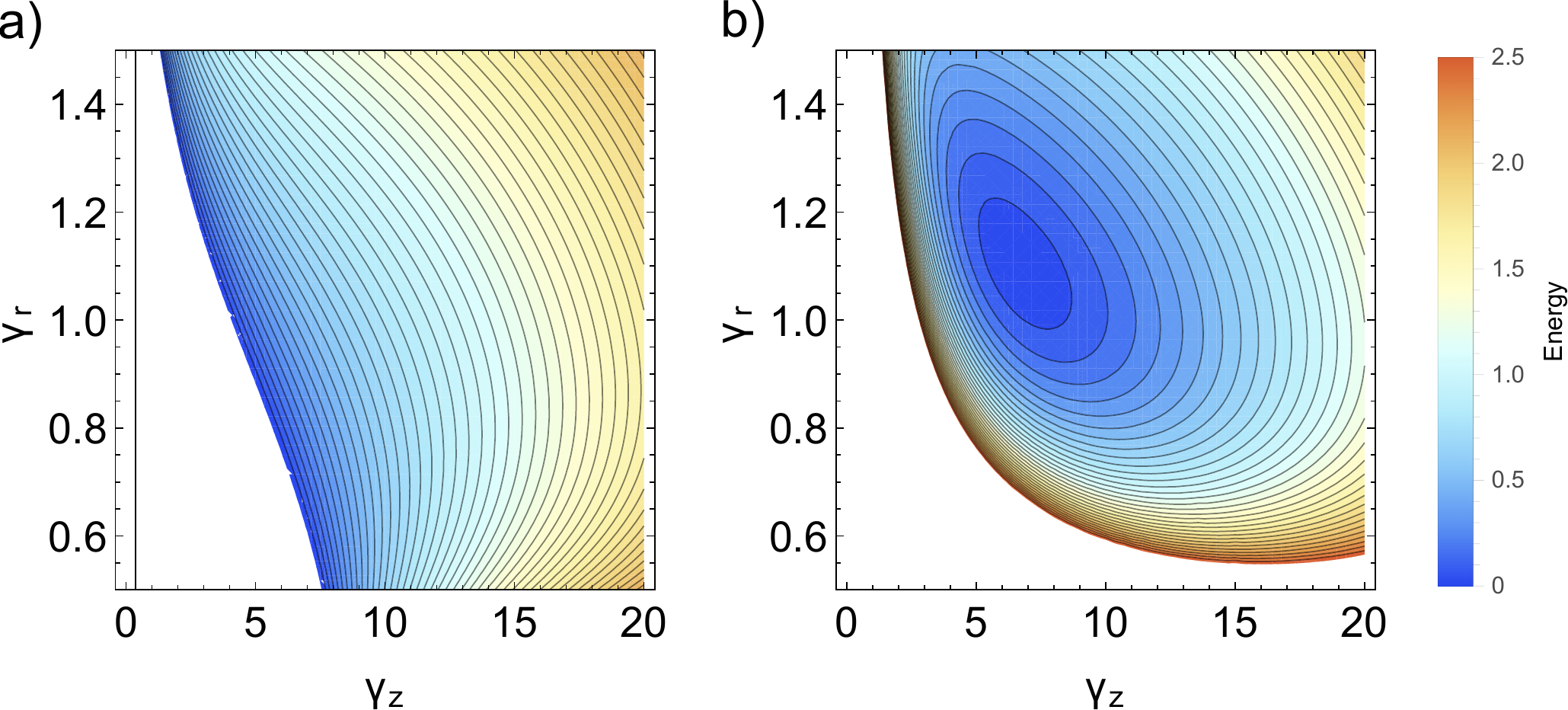}
 	\caption{(Sup) Energy surface plots of equation \ref{eqn:energy2}(sup) against radial and axial width at $a_{s}=-20a_{0}$ for both a) $g_{3}=0$ and b) $\text{Re}[g_{3}]=1.23\times10^{-25}\hbar\si{cm^6s^{-1}}$ (b). Parameters are: \{N=$1.5\times10^{4}$, $\omega_{z}=2\pi\times5.88$  \si{\hertz}, $\omega_{r}=2\pi\times77$  \si{\hertz}\}. For $\text{Re}[g_{3}]=0$ the energy surface tends towards collapse, for non zero $\text{Re}[g_{3}]$ a stable point is shown.}
 	\label{fig:stability}		
 \end{figure}

\begin{figure}[!htp]
	\centering{}	
	\includegraphics[width=1\columnwidth]{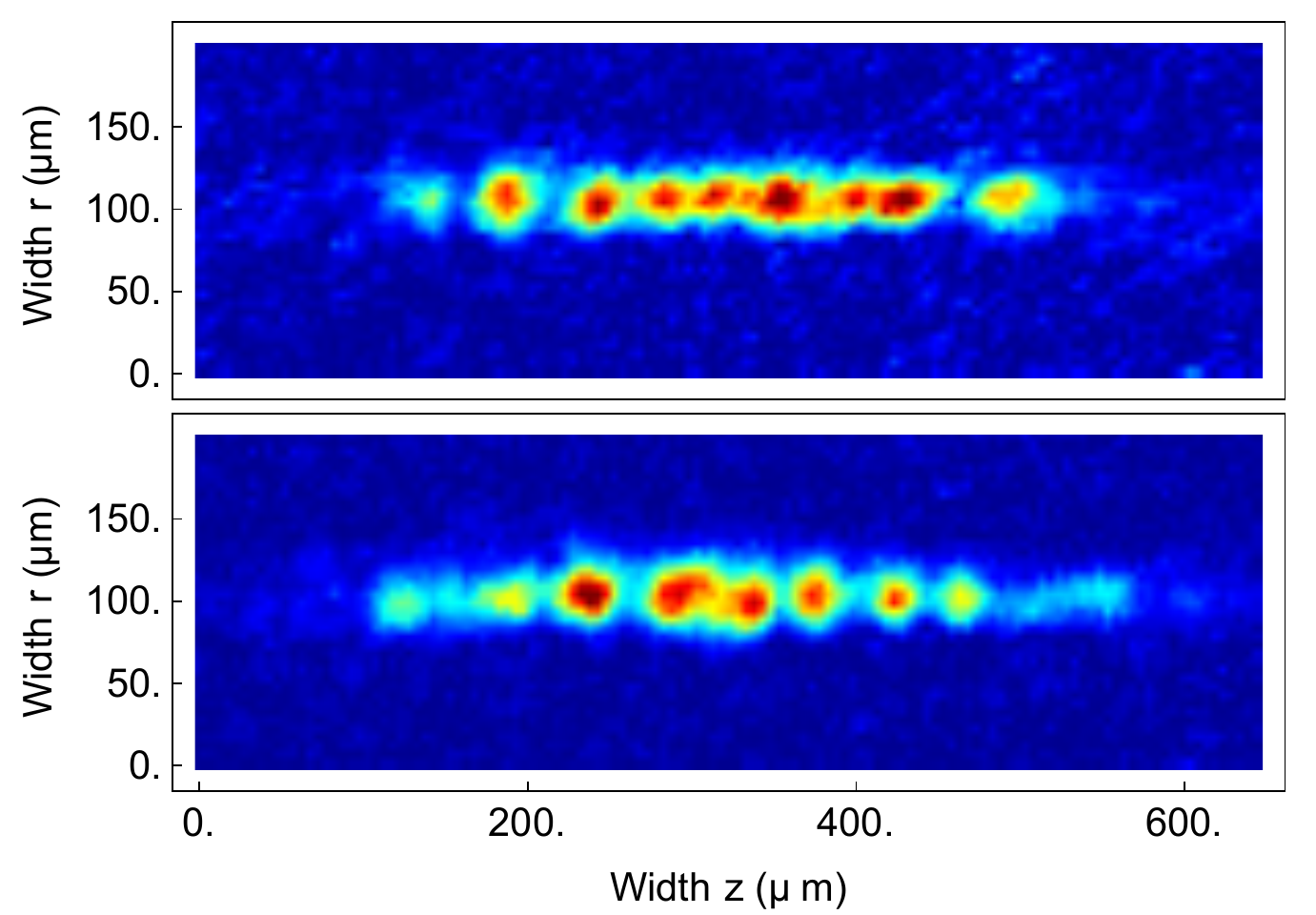}
	\caption{(Sup) In further experiments, soliton trains were produced via modulational instability in a single beam optical waveguide. Total atom number in these images is $1.5\times10^4$, and expansion time from the trap is 20ms. The imaging system is clearly able to resolve multiple solitons.}
	\label{fig:trains}		
\end{figure}

Following a previous method and the same dimensionless parameters\cite{carr_dynamics_2002}, the Gross-Pitaevski energy functional was used to calculate the energy surface with respect to both the axial and radial size of the condensate for the cubic-quintic GP:
\begin{equation}
\label{eqn:energy2}
\epsilon_{{\rm GP}}=
\frac{1}{2\,\gamma_{r}^{2}}
+\frac{\gamma_{r}^2}{2}
+\frac{1}{6\,\gamma_z^2}
+\frac{\pi^2}{24}\lambda^2\,\gamma_z^2
+\frac{\alpha}{3\gamma_{r}^2\,\gamma_z}
+\frac{2 \beta}{135 \pi^{2} \gamma_z^2 \gamma_{r}^4 }
\end{equation}

where $\beta = N^{2} \text{Re}[g_{3}] m /\left( \hbar^{2}  \sigma_{r}^{4}\right)$. In figure \ref{fig:stability}(sup) the energy surface is plotted for $a_{s}=-20a_{0}$ and the system parameters given in the main text. With $\text{Re}[g_{3}]=0$, no stable point is present and the widths would tend towards collapse. However, for $\text{Re}[g_{3}]=1.23\times10^{-25}\hbar\si{cm^6s^{-1}}$ a stable point is clearly present. The inclusion of a positive Re[$g_3$] stabilises the condensate at large negative scattering lengths\cite{akhmediev_bose-einstein_1999}, consistent with experimental observation of breathers in this regime.

\subsection{V. Phase shift arising from ballistic expansion}
\begin{figure}[!htp]
	\centering{}	
	\includegraphics[width=1\columnwidth]{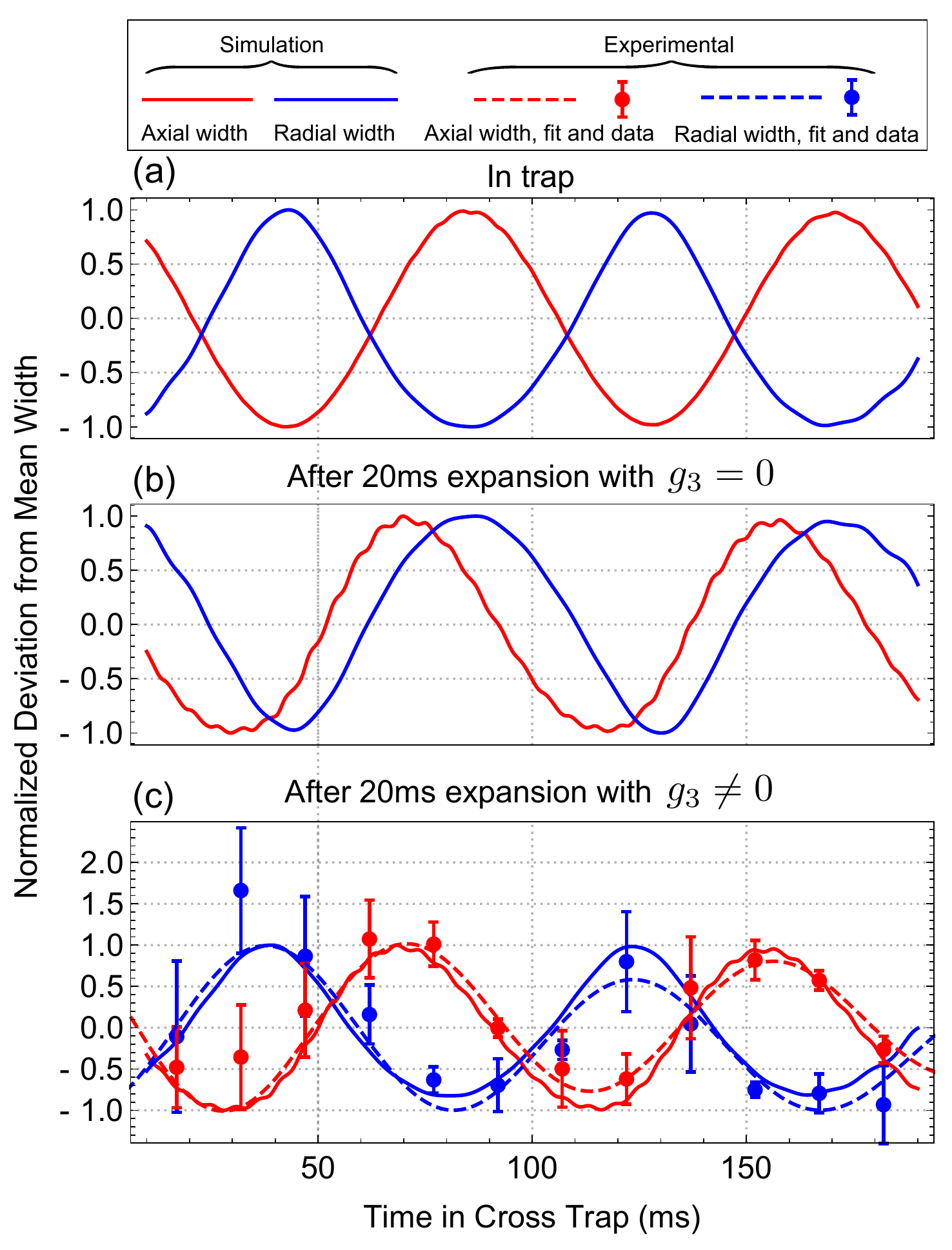}
	\caption{(Sup)Axial and radial widths of a simulated BEC with $N=1.5\times10^4$, undergoing quadrupole oscillations at $-13a_{0}$ inside a cigar shaped trap ($\omega_{z}=2\pi\times5.88$, $\omega_{r}=2\pi\times77$ \si{\hertz}). a) The in trap widths of the condensate showing the out of phase oscillations characteristic of the quadrupole mode. b) The widths of the condensate after undergoing 20ms of expansion with $a_{s}=0$ and Re[$g_3$]=0 at each point in time. c) The expanded widths with Re[$g_3$]$=1.23\times10^{-25}\hbar\si{cm^6s^{-1}}$ plotted with the experimental data}
	\label{fig:phase_shift}		
\end{figure}

As the cloud is initially released from the optical trap at high density, repulsive three body interactions during the dynamics of the expansion change the phase of the measured cloud widths. It is thus important to include $g_{3}$ when simulating ballistic expansion of the cold cloud prior to imaging. Figure \ref{fig:phase_shift}(sup) shows a simulation of the quadrupole oscillation for the same conditions as the experimental data in Figure 2 of the main text. The simulated expansion with $a_{s}=0$ and Re[$g_3$]$=1.23\times10^{-25}\hbar\si{cm^6s^{-1}}$ is plotted with the experimental data in Figure \ref{fig:phase_shift}c. The phase shift between the axial and radial oscillations of $0.25\pi$ agrees with the measured shift of $\left(0.24\pm0.06\right)\pi$. The same expansion with Re[$g_3$]=0 does not agree with the experimental data (Figure \ref{fig:phase_shift}b).\\

\newpage
\bibliographystyle{unsrt}
\bibliography{Breathers_arXiv_main.bbl}

\begin{thebibliography}{10}

\bibitem{ablowitz_solitons_1991}
Mark~J. Ablowitz and P.~A. Clarkson.
\newblock {\em Solitons, {Nonlinear} {Evolution} {Equations} and {Inverse}
  {Scattering}}.
\newblock Cambridge University Press, December 1991.

\bibitem{kh._abdullaev_dynamics_2005}
Fatkhulla Kh.~Abdullaev, Arnaldo Gammal, Anatoly~M. Kamchatnov, and Lauro
  Tomio.
\newblock Dynamics of bright matter wave solitons in a {Bose}-{Einstein}
  condensate.
\newblock {\em International Journal of Modern Physics B}, 19(22):3415--3473,
  September 2005.

\bibitem{matuszewski_stability_2005}
M.~Matuszewski, E.~Infeld, G.~Rowlands, and M.~Trippenbach.
\newblock Stability analysis of three-dimensional breather solitons in a
  {Bose}-{Einstein} condensate.
\newblock {\em Proceedings of the Royal Society of London A: Mathematical,
  Physical and Engineering Sciences}, 461(2063):3561--3574, November 2005.

\bibitem{cardoso_modulation_2010}
W.~B. Cardoso, A.~T. Avelar, and D.~Bazeia.
\newblock Modulation of breathers in cigar-shaped {Bose}-{Einstein}
  condensates.
\newblock {\em Physics Letters A}, 374(26):2640--2645, June 2010.

\bibitem{trombettoni_discrete_2001}
Andrea Trombettoni and Augusto Smerzi.
\newblock Discrete {Solitons} and {Breathers} with {Dilute} {Bose}-{Einstein}
  {Condensates}.
\newblock {\em Physical Review Letters}, 86(11):2353--2356, March 2001.

\bibitem{dunjko_superheated_2014}
Vanja Dunjko and Maxim Olshanii.
\newblock Superheated integrability and multisoliton survival through
  scattering off barriers.
\newblock {\em arXiv:1501.00075 [cond-mat, physics:nlin]}, December 2014.
\newblock arXiv: 1501.00075.

\bibitem{kibler_peregrine_2010}
B.~Kibler, J.~Fatome, C.~Finot, G.~Millot, F.~Dias, G.~Genty, N.~Akhmediev, and
  J.~M. Dudley.
\newblock The {Peregrine} soliton in nonlinear fibre optics.
\newblock {\em Nature Physics}, 6(10):790--795, October 2010.

\bibitem{kosevich_magnetic_1998}
A.~M. Kosevich, V.~V. Gann, A.~I. Zhukov, and V.~P. Voronov.
\newblock Magnetic soliton motion in a nonuniform magnetic field.
\newblock {\em Journal of Experimental and Theoretical Physics},
  87(2):401--407, August 1998.

\bibitem{mcdonald_bright_2014}
G.D. McDonald, C.~C.~N. Kuhn, K.~S. Hardman, S.~Bennetts, P.~J. Everitt, P.~A.
  Altin, J.~E. Debs, J.~D. Close, and N.~P. Robins.
\newblock Bright {Solitonic} {Matter}-{Wave} {Interferometer}.
\newblock {\em Physical Review Letters}, 113(013002), February 2014.

\bibitem{frantzeskakis_dark_2010}
D.~J. Frantzeskakis.
\newblock Dark solitons in atomic {Bose}-{Einstein} condensates: from theory to
  experiments.
\newblock {\em Journal of Physics A: Mathematical and Theoretical},
  43(21):213001, May 2010.

\bibitem{nguyen_collisions_2014}
Jason H.~V. Nguyen, Paul Dyke, De~Luo, Boris~A. Malomed, and Randall~G. Hulet.
\newblock Collisions of matter-wave solitons.
\newblock {\em Nature Physics}, 10(12):918--922, December 2014.

\bibitem{cornish_formation_2006}
Simon~L. Cornish, Sarah~T. Thompson, and Carl~E. Wieman.
\newblock Formation of {Bright} {Matter}-{Wave} {Solitons} during the
  {Collapse} of {Attractive} {Bose}-{Einstein} {Condensates}.
\newblock {\em Physical Review Letters}, 96(17):170401, May 2006.

\bibitem{marchant_controlled_2013}
A.~L. Marchant, T.~P. Billam, T.~P. Wiles, M.~M.~H. Yu, S.~A. Gardiner, and
  S.~L. Cornish.
\newblock Controlled formation and reflection of a bright solitary matter-wave.
\newblock {\em Nature Communications}, 4:1865, May 2013.

\bibitem{kivshar_dynamics_1989}
Yuri~S. Kivshar and Boris~A. Malomed.
\newblock Dynamics of solitons in nearly integrable systems.
\newblock {\em Reviews of Modern Physics}, 61(4):763--915, October 1989.

\bibitem{flach_discrete_2008}
Sergej Flach and Andrey~V. Gorbach.
\newblock Discrete breathers -- {Advances} in theory and applications.
\newblock {\em Physics Reports}, 467(1–3):1--116, October 2008.

\bibitem{mandelik_observation_2003}
D.~Mandelik, H.~S. Eisenberg, Y.~Silberberg, R.~Morandotti, and J.~S.
  Aitchison.
\newblock Observation of {Mutually} {Trapped} {Multiband} {Optical} {Breathers}
  in {Waveguide} {Arrays}.
\newblock {\em Physical Review Letters}, 90(25):253902, June 2003.

\bibitem{peyrard_using_1998}
M.~Peyrard.
\newblock Using {DNA} to probe nonlinear localized excitations?
\newblock {\em EPL (Europhysics Letters)}, 44(3):271, November 1998.

\bibitem{Akhmediev}
Private communication with Prof. Nail Akhmediev, Optical Sciences Group, ANU.

\bibitem{pollack_collective_2010}
S.~E. Pollack, D.~Dries, R.~G. Hulet, K.~M.~F. Magalhaes, E.~A.~L. Henn,
  E.~R.~F. Ramos, M.~A. Caracanhas, and V.~S. Bagnato.
\newblock Collective excitation of a {Bose}-{Einstein} condensate by modulation
  of the atomic scattering length.
\newblock {\em Physical Review A}, 81(5):053627, May 2010.

\bibitem{altin_collapse_2011}
P.~A. Altin, G.~R. Dennis, G.~D. McDonald, D.~Doring, J.~E. Debs, J.~D. Close,
  C.~M. Savage, and N.~P. Robins.
\newblock Collapse and three-body loss in a $^{85}${Rb} {Bose}-{Einstein}
  condensate.
\newblock {\em Physical Review A}, 84(3):033632, September 2011.

\bibitem{kohler_three-body_2002}
Thorsten K\"ohler.
\newblock Three-{Body} {Problem} in a {Dilute} {Bose}-{Einstein} {Condensate}.
\newblock {\em Physical Review Letters}, 89(21):210404, November 2002.

\bibitem{al-jibbouri_geometric_2013}
Hamid Al-Jibbouri, Ivana Vidanovic, Antun Balaz, and Axel Pelster.
\newblock Geometric resonances in {Bose}-{Einstein} condensates with two- and
  three-body interactions.
\newblock {\em Journal of Physics B: Atomic, Molecular and Optical Physics},
  46(6):065303, March 2013.

\bibitem{hao-cai_boseeinstein_2010}
Li~Hao-Cai, Chen Hai-Jun, and Xue Ju-Kui.
\newblock Bose-{Einstein} {Condensates} with {Two}- and {Three}-{Body}
  {Interactions} in an {Anharmonic} {Trap} at {Finite} {Temperature}.
\newblock {\em Chinese Physics Letters}, 27(3):030304, March 2010.

\bibitem{gammal_critical_2001}
A.~Gammal, T.~Frederico, and Lauro Tomio.
\newblock Critical number of atoms for attractive {Bose}-{Einstein} condensates
  with cylindrically symmetrical traps.
\newblock {\em Physical Review A}, 64(5):055602, October 2001.

\bibitem{pieri_derivation_2003}
P.~Pieri and G.~C. Strinati.
\newblock Derivation of the {Gross}-{Pitaevskii} {Equation} for {Condensed}
  {Bosons} from the {Bogoliubov}-de {Gennes} {Equations} for {Superfluid}
  {Fermions}.
\newblock {\em Physical Review Letters}, 91(3):030401, July 2003.

\bibitem{PhysRevLett.30.1262}
M.~J. Ablowitz, D.~J. Kaup, A.~C. Newell, and H.~Segur.
\newblock Method for solving the {Sine}-{Gordon} equation.
\newblock {\em Phys. Rev. Lett.}, 30:1262--1264, Jun 1973.

\bibitem{donley_dynamics_2001}
Elizabeth~A. Donley, Neil~R. Claussen, Simon~L. Cornish, Jacob~L. Roberts,
  Eric~A. Cornell, and Carl~E. Wieman.
\newblock Dynamics of collapsing and exploding {Bose}-{Einstein} condensates.
\newblock {\em Nature}, 412(6844):295--299, July 2001.

\bibitem{snoke2015}
Y.~{Sun}, Y.~{Yoon}, M.~{Steger}, G.~{Liu}, L.~N. {Pfeiffer}, K.~{West}, D.~W.
  {Snoke}, and K.~A. {Nelson}.
\newblock {Polaritons are Not Weakly Interacting: Direct Measurement of the
  Polariton-Polariton Interaction Strength}.
\newblock {\em ArXiv e-prints}, August 2015.

\bibitem{kozuma_coherent_1999}
M.~Kozuma, L.~Deng, E.~W. Hagley, J.~Wen, R.~Lutwak, K.~Helmerson, S.~L.
  Rolston, and W.~D. Phillips.
\newblock Coherent {Splitting} of {Bose}-{Einstein} {Condensed} {Atoms} with
  {Optically} {Induced} {Bragg} {Diffraction}.
\newblock {\em Physical Review Letters}, 82(5):871--875, February 1999.

\bibitem{kuhn_bose-condensed_2014}
C.~C.~N. Kuhn, G.~D. McDonald, K.~S. Hardman, S.~Bennetts, P.~J. Everitt, P.~A.
  Altin, J.~E. Debs, J.~D. Close, and N.~P. Robins.
\newblock A {Bose}-condensed, simultaneous dual-species {Mach}-{Zehnder} atom
  interferometer.
\newblock {\em New Journal of Physics}, 16(7):073035, July 2014.

\bibitem{ronen_bogoliubov_2006}
Shai Ronen, Daniele C.~E. Bortolotti, and John~L. Bohn.
\newblock Bogoliubov modes of a dipolar condensate in a cylindrical trap.
\newblock {\em Physical Review A}, 74(1):013623, July 2006.

\bibitem{1367-2630-16-12-122001}
Romain Dubessy, Camilla~De Rossi, Thomas Badr, Laurent Longchambon, and Helene
  Perrin.
\newblock Imaging the collective excitations of an ultracold gas using
  statistical correlations.
\newblock {\em New Journal of Physics}, 16(12):122001, December 2014.

\bibitem{perez-garcia_low_1996}
Victor~M. Perez-Garcia, H.~Michinel, J.~I. Cirac, M.~Lewenstein, and P.~Zoller.
\newblock Low {Energy} {Excitations} of a {Bose}-{Einstein} {Condensate}: {A}
  {Time}-{Dependent} {Variational} {Analysis}.
\newblock {\em Physical Review Letters}, 77(27):5320--5323, December 1996.

\bibitem{bao_three-dimensional_2004}
Weizhu Bao, D.~Jaksch, and P.~A. Markowich.
\newblock Three-dimensional simulation of jet formation in collapsing
  condensates.
\newblock {\em Journal of Physics B: Atomic, Molecular and Optical Physics},
  37(2):329, January 2004.

\bibitem{saito_mean-field_2002}
Hiroki Saito and Masahito Ueda.
\newblock Mean-field analysis of collapsing and exploding {Bose}-{Einstein}
  condensates.
\newblock {\em Physical Review A}, 65(3):033624, March 2002.

\bibitem{roberts_magnetic_2000}
J.~L. Roberts, N.~R. Claussen, S.~L. Cornish, and C.~E. Wieman.
\newblock Magnetic {Field} {Dependence} of {Ultracold} {Inelastic} {Collisions}
  near a {Feshbach} {Resonance}.
\newblock {\em Physical Review Letters}, 85(4):728--731, July 2000.

\bibitem{stringari_collective_1996}
S.~Stringari.
\newblock Collective {Excitations} of a {Trapped} {Bose}-{Condensed} {Gas}.
\newblock {\em Physical Review Letters}, 77(12):2360--2363, September 1996.

\bibitem{Hulet2015}
Private communication with Prof. Randy Hulet, Physics and Astronomy Department,
  Rice University.

\bibitem{RevModPhys.82.1225}
Cheng Chin, Rudolf Grimm, Paul Julienne, and Eite Tiesinga.
\newblock Feshbach resonances in ultracold gases.
\newblock {\em Rev. Mod. Phys.}, 82:1225--1286, Apr 2010.

\bibitem{savage_bose-einstein_2003}
C.~M. Savage, N.~P. Robins, and J.~J. Hope.
\newblock Bose-{Einstein} condensate collapse: {A} comparison between theory
  and experiment.
\newblock {\em Physical Review A}, 67(1):014304, January 2003.

\bibitem{carr_dynamics_2002}
L.~D. Carr and Y.~Castin.
\newblock Dynamics of a matter-wave bright soliton in an expulsive potential.
\newblock {\em Physical Review A}, 66(6):063602, December 2002.

\bibitem{akhmediev_bose-einstein_1999}
N.~Akhmediev, M.~P. Das, and A.~V. Vagov.
\newblock Bose-{Einstein} condensation of atoms with attractive interaction.
\newblock {\em International Journal of Modern Physics B}, 13(05n06):625--631,
  March 1999.

\end{thebibliography}

\end{document}